\documentclass[10pt]{article}
\usepackage[usenames,dvipsnames]{xcolor}
\usepackage[OE]{express}

\begin{document}

\title{Photonic temporal-mode multiplexing by quantum frequency
  conversion in a dichroic-finesse cavity}

\author{Dileep V. Reddy\authormark{1,*} and Michael G. Raymer\authormark{1}}

\address{\authormark{1}Oregon Center for Optical, Molecular, and Quantum Science, and Department of Physics, 1274 University of Oregon, Eugene, OR 97403, USA}

\email{\authormark{*}dileep@uoregon.edu} 

\begin{abstract*}
  Photonic temporal modes (TMs) form a field-orthogonal,
  continuous-variable degree of freedom that is in principle infinite
  dimensional, and create a promising resource for quantum information
  science and technology. The ideal quantum pulse gate (QPG) is a
  device that multiplexes and demultiplexes temporally orthogonal
  optical pulses that have the same carrier frequency, spatial mode,
  and polarization. The QPG is the chief enabling technology for usage
  of orthogonal temporal modes as a basis for high-dimensional quantum
  information storage and processing. The greatest hurdle for QPG
  implementation using nonlinear-optical, parametric processes with
  time-varying pump or control fields is the limitation on achievable
  temporal mode selectivity, defined as perfect TM discrimination
  combined with unity efficiency. We propose the use of pulsed
  nonlinear frequency conversion in an optical cavity having greatly
  different finesses for different frequencies to implement a nearly
  perfectly TM-selective QPG in a low-loss integrated-optics platform.
\end{abstract*}

\bibliographystyle{osajnl}

An important goal in quantum information science and technology is
complete control of photonic states \cite{ptd12}. Beyond the
polarization and transverse spatial degrees of freedom, the
time-frequency degree of freedom is largely an untapped quantum
resource \cite{Humphreys:2014ce,nunn13,Wright,Davis:17}. Orthogonal
temporal modes (TMs) are defined by the complex longitudinal
wave-packet shape functions of pulsed modes of
light\cite{Smith:2007njp}. They form a field-orthogonal,
continuous-variable degree of freedom that is in principle infinite
dimensional, and create a promising resource for quantum information
science\cite{Brecht:15prx}. To fully exploit their use in a quantum
network requires the ability to unitarily demultiplex arbitrary TM
components from a light beam with near-unity efficiency and mode
discrimination (i.e., no crosstalk). A device capable of such
operations is known as a quantum pulse gate
(QPG)\cite{EcksteinA:2011vg,Brecht:2011hz}. The ideal QPG must satisfy
two conditions: (a) It must fully separate the desired TM component
from the others without loss of photons, and (b) it must avoid
contamination from orthogonal TM components in the ``wrong'' TM
channels. When both of these are met, the QPG is said to have unit
selectivity \cite{Reddy:2013ip}.

There exists a fundamental limit to selectivity of QPGs based on
traveling-wave interactions in media with simple dispersion profiles,
which enforces a trade-off between the two aforementioned
conditions\cite{Reddy:2013ip,Ansari:16ax,Manurkar:16,Reddy:17}. The
best performing QPG proposed to date is based on temporal-mode
interferometry (TMI) that performs pulsed, cascaded frequency
conversion with multiple passes through standard dispersive nonlinear
optical media (three-wave mixing in crystals or four-wave mixing in
fiber) \cite{Reddy:2014bt,Reddy:15pra,Kobayashi:17ax}. The technique
is a close relative of Ramsey interference of photons in a
frequency-converting interferometer \cite{Gaeta:15ax,Reddy:15pra}. In
TMI, TMs from all participating carrier-frequency bands coherently
reinteract at every stage, resulting in a selectivity enhancement that
overcomes the single-stage maximum \cite{Quesada:16}. The gain in
selectivity is very significant for even two-stage schemes, and is
predicted to improve asymptotically with the number of
stages\cite{Reddy:15pra}. While TMI has been shown to operate as
predicted, it presents practical difficulties due to coupling losses
and engineering/manufacturing constraints for integration.

A QPG that can multiplex and demultiplex field-orthogonal optical
pulses is closely related to devices studied for coherent optical code
multiple access (OCDMA) employing second-harmonic generation of
phase-structured ultrafast pulses \cite{Zheng:01,Zheng:02}. Like the
other implementations of coherent demultiplexers mentioned above, the
TM selectivity of this scheme is limited by traveling-wave phenomena
during nonlinear frequency conversion \cite{Zheng:02}. In a different
research arena, optical-cavity-enhanced atomic-ensemble or solid-state
quantum memories are known theoretically to have TM-selective
qualities for coherent optical storage \cite{Cirac:04ax,
  Bao:12,Nunn:16ax}. Finally, great technical advances have been made
in design and fabrication of nonlinear-optical micro-ring resonators,
and these have been employed for frequency conversion between telecom
and visible bands via sum-frequency generation
\cite{Strekalov,Li,Guo,kumarcav}.

By combining insights from all the diverse areas discussed above, we
have arrived at a means of using an optical micro-cavity with a large
difference in finesse for two frequency bands participating in
nonlinear frequency conversion by sum-frequency generation (SFG), to
mimic the TM-selective behavior predicted for cavity-based atomic
quantum memories. This all-optical ``dichroic-finesse cavity'' scheme
provides a simple, realistic way to create a near-ideal add/drop
(multiplexer/demultiplexer) device in a low-loss integrated-optics
platform for use in quantum optical networks. By passing the control
and signal pulses through the same frequency-converting medium many
times, the device effectively performs TMI with a near-infinite number
of stages, which explains its high TM selectivity. The proposed system
operates without the need for atomic vapors or doped crystals. The
weak signal pulse may be in a single-photon state (or any low-number
Fock state), or in any other quantum state, such as squeezed
vacuum. It can be temporally reshaped during the read-in and read-out
process. The proposed scheme, while challenging to construct, relies
only on already proven technology.

\begin{figure}[bht]
  \centering
  \includegraphics[width=\linewidth]{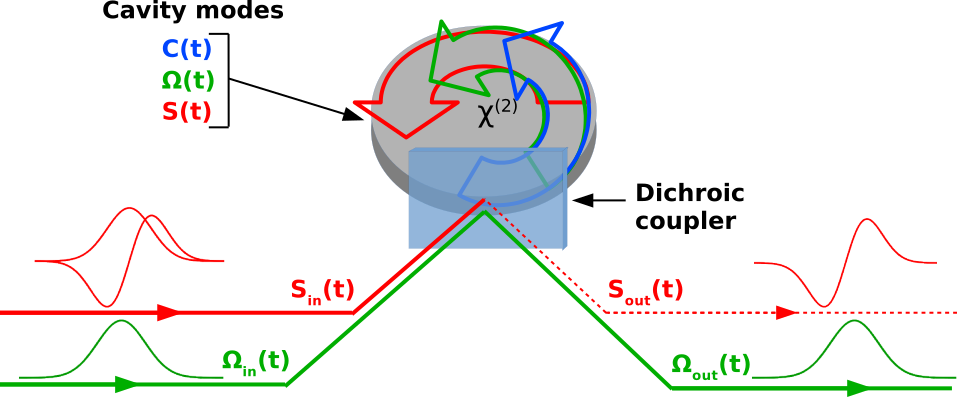}
  \caption{Schematic of dichroic-finesse cavity filled with
    second-order nonlinear optical material. The converted field
    $C(t)$ is not shown exiting the cavity, as this occurs on much
    longer time scales.}
  \label{fig01}
\end{figure}

Figure \ref{fig01} shows a schematic diagram of the proposed
micro-ring cavity system. $S(t)$ and $C(t)$ are the resonant cavity
mode amplitudes of the two frequency bands that will interact via the
optical nonlinearity of the medium, which uniformly fills the
cavity. The intracavity control field $\Omega (t)$ can be a single
strong, coherent laser pulse if the process is utilizing three-wave
mixing, or a combination (product) of two pulses if four-wave mixing
is used. For definiteness, we will consider sum-frequency generation
by three-wave mixing, but all results apply to four-wave mixing as
well. For sum-frequency generation the frequencies are related by
${{\omega }_{s}}+{{\omega }_{\Omega }}={{\omega }_{c}}$. For
convenience we refer to ${{\omega }_{s}}$ as ``red'' and ${{\omega
  }_{c}}$ as ``blue.''

The theoretical model presented below predicts that for a given
temporal shape of the control field $\Omega (t)$ inside the cavity,
only a single temporal mode of the incoming (red) signal field, called
the target mode, will be frequency up-converted, creating narrowband
(blue) light that is trapped in the cavity in the $C(t)$ mode. All TMs
temporally orthogonal to the target mode will be transmitted into the
${{S}_{\text{\text out}}}(t)$ beam with the original (red) carrier
frequency. The trapped blue light can subsequently be left to leak
slowly from the cavity at later times, or it can be rapidly read out
(ejected) from the cavity by applying a subsequent control pulse,
which converts it back to red.

The temporal widths of the control field and the signal input
${{S}_{\text{\text in}}}(t)$ must be much longer than the cavity round-trip
time, but much shorter than the cavity lifetime for the $C(t)$
mode. This leads to negligible leakage of the converted $C(t)$
amplitude from the cavity during the process. The $C(t)$ mode then
becomes analogous to a coherent spin wave (for example) in a cold
atomic ensemble. The cavity-coupling coefficients are assumed to put
$S(t)$ in the ``bad-cavity'' limit. This finesse differential across the
frequency bands is the key feature necessary to break the interaction
symmetry and yield efficient TM multiplexing.

The solutions to the coupled-mode equations of motion can be expressed
as linear integral scattering relations between input and output
temporal modes using Green's functions, which are functions of both an
input-mode time argument and an output-mode time argument. For the
process to be mode discriminatory, the Green's functions should be
separable in their time arguments, which is impossible for
time-stationary processes. A key requirement for achieving Green's
function separability in previous (cavityless) approaches has been a
large difference in the group velocities between the various frequency
bands\cite{Reddy:2013ip,Reddy:2014bt,Reddy:15pra,Reddy:17}. This is
required because orthogonal TMs can share very similar (even
identical) temporal features in local time slices. For the QPG to
perform different transformations on these two TMs, the full global
mode structure needs to be surveyed by the device, as the effect
(depletion/enhancement/phase-shift) on any given time slice should
depend on features in all other time slices. Differing group
velocities cause pulsed modes field amplitudes from different time
slices to convect through each other, providing an effective means of
carrying local mode information across different time slices. TMI
operates by causing convecting pulses to overlap in spacetime over
multiple stages, with the interaction being semi-perturbative during
each stage. This avoids coherent-propagation ringing effects
\cite{Burnham1969} induced by cascaded second-order nonlinearity
\cite{Reddy:2013ip,Zheng:01,Zheng:02}, and ensures Green's function
separability even at high conversion efficiencies\cite{Quesada:16}.

Our proposed scheme ensures inter-pulse convection by confining the
TMs of one of the bands in physical space as the other TMs pass
through it. Our design using a large difference in finesse across the
frequency bands works for controls and signals with arbitrary relative
group velocities, which is another advantage over (cavityless)
traveling-wave QPG implementations.

We analyze the case of a nonlinear waveguide forming a resonant ring
cavity, assuming frequencies that are phase matched for the control
field $\Omega (t)$, signal ${{S}_{\text{\text in}}}(t)$, and converted
$C(t)$. The key to is to have the cavity input-output coupling be
frequency dependent, while still requiring that both modes have high
finesse. With a very long, smooth input signal pulse, this allows the
bad-cavity limit (only) for the $S$-field, meaning it tends to leak
from the cavity relatively rapidly. Concurrently, we assume the cavity
has much higher finesse for the $C$-field than for the $S$
field. Then, we can frequency convert the short, ``red'' input pulse
${{S}_{\text{\text in}}}(t)$ into a long-lived, resonant cavity mode at the
``blue'' frequency ${{\omega }_{c}}$ trapped within the cavity before
it slowly leaks exponentially from the cavity.

Within the adopted parameter ranges, the system is well described
using the standard input-output theory of Collett and Gardiner
\cite{Collett}. The approximations leading to this formalism require
very weak coupling of the cavity modes to external freely propagating
modes, and spectral widths of all signals significantly narrower than
the free-spectral range of the cavity \cite{Raymercav}.

The weak quantum signal fields within the cavity are represented by
annihilation operators $S(t)$ for the ``red'' input field and $C(t)$
for the ``blue'' converted field, and satisfy commutators
$[C(t),{{C}^{\dagger }}(t)]=1$, $[S(t),{{S}^{\dagger }}(t)]=1$. The
input fields immediately outside of the coupling mirror are
${{S}_{\text{\text in}}}(t),\ {{C}_{\text{\text in}}}(t)$, which satisfy
$[A_j(t),A^\dagger_k(t')]=\delta_{jk}\delta(t-t')$, and the outgoing
fields are ${{S}_{\text{\text out}}}(t),\,{{C}_{\text{\text out}}}(t)$.

We take $\Omega (t)$ to be the intracavity control field in an
auxiliary mode, which in the ``bad-cavity'' limit is simply proportional
to an incident field ${{\Omega }_{\text in}}(t)$. We absorb the square-root
of the control field energy into a nonlinear interaction parameter
$\alpha $ such that $\Omega (t)$ is square-normalized to one. We
assume both signal fields are exactly resonant with their cavity modes
and there is no phase mismatch for the SFG process. Then the equations
of motion within the cavity are \cite{kumarcav2,Li}: 

\begin{align}
  & {{\partial }_{t}}S(t)=i\alpha{{\Omega }^{*}}(t)C(t)-{{{\tilde{\gamma }}}_{s}}S(t)+\sqrt{2{{\gamma }_{s}}}{{S}_{\text{\text in}}}(t), \\ 
 & {{\partial }_{t}}C(t)=i\alpha \Omega (t)S(t)-{{{\tilde{\gamma }}}_{c}}C(t)+\sqrt{2{{\gamma }_{c}}}{{C}_{\text{\text in}}}(t).
\end{align}

\noindent where ${{\tilde{\gamma }}_{s}}={{\gamma }_{s}}+{{\kappa
  }_{s}},\ {{\tilde{\gamma }}_{c}}={{\gamma }_{c}}+{{\kappa
  }_{c}}$. The (real) damping rates ${{\gamma }_{j}},{{\kappa
  }_{j}}\,(j=s,c)$ correspond to unitary decay from the cavity mode to
the external modes, and nonunitary decay to internal dissipative loss,
respectively. For simplicity, as in \cite{Nunn:16ax}, we omit the
Langevin noise operators associated with the dissipative loss, as they
do not contribute to measured signal intensities. The input-output
relations are (with a chosen phase convention):

\begin{equation}
  {{S}_{\text{\text out}}}(t)=-{{S}_{\text{\text in}}}(t)+\sqrt{2{{\gamma }_{s}}}S(t),\quad {{C}_{\text{\text out}}}(t)=-{{C}_{\text{\text in}}}(t)+\sqrt{2{{\gamma }_{c}}}C(t).
\end{equation}
	 
\noindent In the following, we assume there is no external input to the $C$
mode, so ${{C}_{\text{\text in}}}(t)$ is omitted.

Equations 2 and 3 are linear in field operators (although nonlinear
with respect to the control field, here an undepleted coherent
state). Therefore they can describe the Heisenberg-picture operator
dynamics of any quantum state of light. In the case that only a single
signal photon is present throughout the system, the variables can be
interpreted as Schrodinger-picture state amplitudes
\cite{shen,Mirza:13}.

The first crucial assumption is that the cavity out-coupling rate
${{\gamma }_{s}}$ for the input channel is large compared to the rate
at which all the fields vary - set by ${{\tilde{\gamma
  }}_{s}},{{\tilde{\gamma }}_{c}}$ and $\alpha $, so we can apply the
``bad-cavity'' limit to $S(t)$. By setting ${{\partial }_{t}}S(t)\to
0$, we get

\begin{align}
  & S(t)=i(\alpha /{{{\tilde{\gamma }}}_{s}}){{\Omega }^{*}}(t)C(t)+\sqrt{2{{\gamma }_{s}}/\tilde{\gamma }_{s}^{2}}{{S}_{\text{\text in}}}(t) \\ 
 & {{\partial }_{t}}C(t)=\left[ -{{f}_{s}}|\Omega (t){{|}^{2}}-{{{\tilde{\gamma }}}_{c}} \right]C(t)+i{{g}_{s}}\Omega (t){{S}_{\text{\text in}}}(t).\label{eq5}
\end{align}

\noindent where ${{f}_{s}}={{\alpha }^{2}}/{{\tilde{\gamma
  }}_{s}},\ \ \ {{g}_{s}}=\alpha \sqrt{2{{\gamma }_{s}}/\tilde{\gamma
  }_{s}^{2}}$.

The second crucial assumption is that the cavity has very high finesse
(is a very ``good'' cavity) for the $C$-band (${{\bar{\gamma
  }}_{c}}\approx 0$), and the entire process takes place well before
any amplitude from $C(t)$ has leaked out. Dropping the
${{\tilde{\gamma }}_{c}}C(t)$ term, the solution to eq. \ref{eq5} is

\begin{equation}
  C(t)=i{{g}_{s}}{{e}^{-{{f}_{s}}\epsilon (t)}}\int\limits_{-\infty }^{t}{{{e}^{{{f}_{s}}\epsilon (t')}}\Omega (t'){{S}_{\text{\text in}}}(t')dt'},
  \end{equation}

\noindent where $\epsilon (t)=\int_{-\infty }^{t}{\,|\Omega
  (t''){{|}^{2}}dt''}$. The SFG-cavity mode amplitude at the end of
the process $C(\infty )$ equals zero for any ${{S}_{\text{\text in}}}(t')$
that is orthogonal to ${{e}^{{{f}_{s}}\epsilon (t')}}{{\Omega
  }^{*}}(t')$. The function ${{e}^{{{f}_{s}}\epsilon (t')}}{{\Omega
  }^{*}}(t')$ is thus the optimal TM for storage in this
cavity. Hence, the process is perfectly temporal-mode selective,
within the approximations made here. The perfect discrimination arises
from the fact that the Green's function appearing in the integral for
$C(\infty )$ is ${{e}^{-{{f}_{s}}\epsilon
    (t)}}{{e}^{{{f}_{s}}\epsilon (t')}}$, which is separable in the
input and output variables $t,t'$. Define the optimal input TM as:

\begin{equation}
  {{S}_{\text in,opt}}(t)=N{{\Omega }^{*}}(t)\exp \left[
    {{f}_{s}}\int\limits_{-\infty }^{t}{{{\left| \Omega (t')
          \right|}^{2}}dt'} \right],
  \end{equation}

\noindent where $N=\sqrt{2{{f}_{s}}/({{e}^{2{{f}_{s}}}}-1)}$ ensures
that the square of ${{S}_{\text in,opt}}(t)$ integrates to 1.

This dichroic-finesse cavity scheme is not only highly TM
discriminatory, but is also highly efficient, under the assumption
that there are negligible internal dissipative losses, that is
${{\tilde{\gamma }}_{s}}={{\gamma }_{s}}$. In this case, the
efficiency is unity if the output field ${{S}_{\text out}}(t)$ is zero. In
the case that the input is given by eq. 6, the total unconverted
signal energy (photon number) behaves as $W_{\text out}\to \exp(-2f_s)$,
the trend toward zero being achieved with increasing ${{f}_{s}}$. This
prediction is valid only up to a certain value of control field
strength, beyond which the system is driven out of the bad-cavity
regime and the conversion efficiency degrades, as we discuss below.

To verify the scheme operates as a high-selectivity quantum pulse
gate, we solve the more accurate eqs. 1 and 2 numerically. Unless
stated otherwise, the control field is taken to be Gaussian,
$\Omega(t)={(2/\pi )}^{1/4}\exp [-(t-3)^2]$. Time units are relative
to the duration of this dimensionless control pulse. The first goal is
to show that the optimal input pulse, designed according to eq. 7,
leads to efficient transfer of incoming energy into the
frequency-converted cavity mode . (We continue to assume negligible
internal cavity loss.)

Figure 2 shows simulation results in the dichroic-finesse cavity
scenario, for the case ${{\gamma }_{s}}=10.1$ and ${{\gamma
  }_{c}}=0.010$, or dimensionless cavity lifetimes $1/{{\gamma
  }_{s}}=0.09,\ 1/{{\gamma }_{c}}=99.99$. The value of nonlinear
coupling, dependent on control pulse energy, is optimized to be
$\alpha =5.5$. As expected, larger values begin to drive the system
out of the bad-cavity regime and worsen the conversion efficiency (not
shown). The integrated signal input energy equals 1.

Figures 2a and 2b show the result for the case that the input (``red'')
signal shape ${{S}_{\text in,Gaussian}}$ is identical in shape to the
control pulse $\Omega $, not the optimal case. The ``red'' cavity mode
$S$ reaches a maximum of about 0.2 before rapidly decaying. The
converted ``blue'' cavity mode amplitude, plotted as $-iC$, reaches a
value 0.8 before it begins a slow exponential decay into the output
channel ${{C}_{\text out}}$. The ``red'' output channel ${{S}_{\text out}}$ shows
significant leakage and thus poor storage efficiency. The unconverted
signal energy ${{W}_{\text out}}$equals 0.36 in this case with a
non-optimized input pulse shape.

\begin{figure}[bht]
  \centering
  \includegraphics[width=\linewidth]{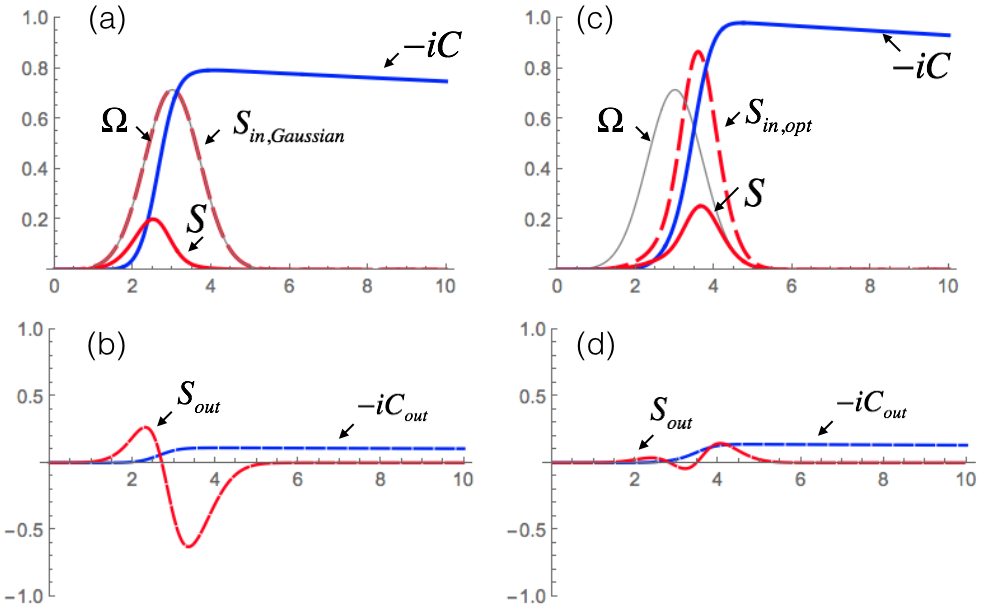}
  \caption{Numerical simulations of amplitude versus time for (a, b)
    Gaussian signal input and (c, d) the optimal input temporal
    mode. The input signal $S_{\text in}$ and the control pulse
    $\Omega$ are multiplied by 0.8 for convenient plotting. Parameters
    for both cases: $\alpha = 5.5, \gamma_s= 10.1, \gamma_c = 0.01$.}
  \label{fig02}
\end{figure}

Figures 2c and 2d show the case that the input signal shape is given
by eq. 6, which is predicted to be optimal. In this case the
unconverted signal energy ${{W}_{\text out}}$ equals 0.016, meaning less
than 2\% of the incoming pulse is not initially trapped in the
frequency-converted cavity mode. Correspondingly, the trapped ``blue''
mode amplitude reaches a value near 1.0 before it begins a slow
exponential decay. This means that a properly designed input pulse can
achieve high storage efficiency, analogously to results found in
atomic-based quantum memories \cite{Nunn:16ax}. 

Any TM orthogonal to the optimal mode given, by eq. 7, is
predicted to pass through the cavity system and not frequency
convert. Figure 3 shows numerical solutions of eqs. 1 and 2 for two
such modes, and, indeed the conversion efficiency is very small for
each. Orthogonal modes, denoted mode 1, mode 2, and so on, are
constructed numerically using a Gram-Schmidt procedure starting from
the optimal mode used earlier in Fig. 2.

Figure 3a shows as the dashed curve the ``red'' input mode 1, which
resembles a Hermite-Gaussian-1 function. The converted ``blue'' cavity
mode amplitude $-iC$ reaches a value $-0.7$ then rapidly returns to a
small value around $-0.1$ before beginning a slow exponential decay into
the output channel ${{C}_{\text out}}$. The unconverted ``red'' output channel
${{S}_{\text out}}$ in Fig. 3b shows large leakage. The unconverted signal
energy ${{W}_{\text out}}$ equals 0.98 in this case, that is it remains
nearly completely unconverted. Figure 3c shows as the dashed curve the
``red'' input mode 2, which resembles a Hermite-Gaussian-2 function. The
converted ``blue'' cavity mode amplitude $-iC$ oscillates and rapidly
returns to a near-zero value, and the ``red'' output channel
${{S}_{\text out}}$ in Fig. 3d shows large leakage. The unconverted signal
energy ${{W}_{\text out}}$ equals 0.99 in this case, again consistent with
expectations.

\begin{figure}[bht]
  \centering
  \includegraphics[width=\linewidth]{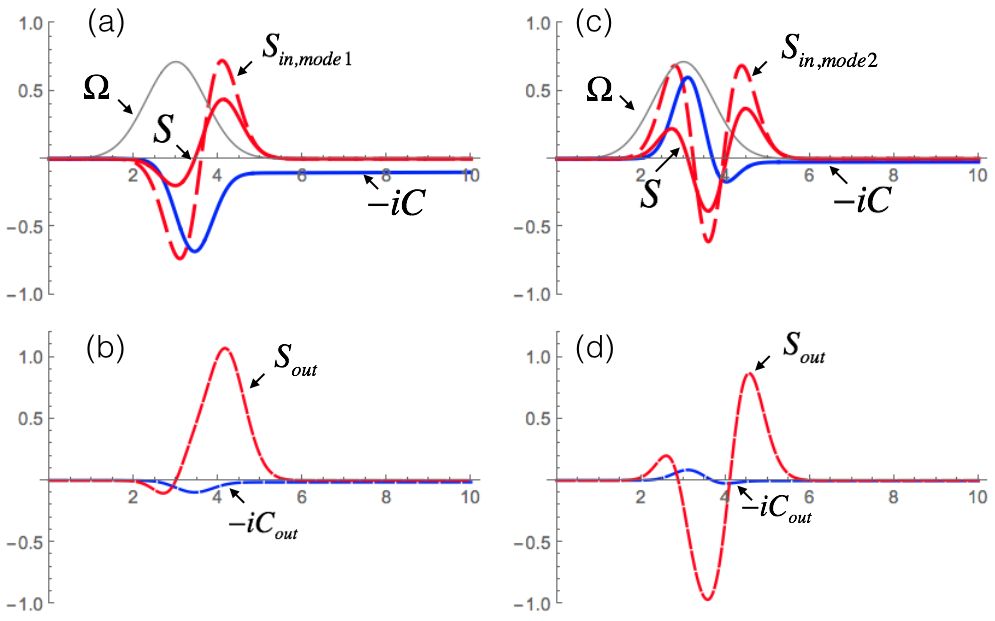}
  \caption{Numerical simulations of amplitude versus time for two
    temporal modes that are orthogonal to the optimum TM used in
    Fig. 2. Both remain nearly completely unconverted. Same parameters
    and plotting as in Fig. 2.}
  \label{fig03}
\end{figure}

The simulations support our proposal for an all-optical cavity-based
demultiplexer: we have shown that given a Gaussian control pulse,
there is one optimal signal TM that can be frequency converted
efficiently and stored for a time in the cavity, while any mode that
is temporally orthogonal to the optimal mode is not frequency
converted and passes through the system. The contrast between energy
conversion and nonconversion is about 50:1 for the parameters and
pulses used, far better than any single-stage QPG based on
traveling-wave SFG.

From a different perspective, if we choose any particular targeted
``red'' signal input TM that we wish to convert to ``blue'' and store in
the dichroic-finesse cavity, we can design the control field that
optimizes its conversion and trapping, while not converting any
orthogonal signal TM. The condition that ensures near-100\% conversion
of the ``red'' input pulse is ${{S}_{\text{\text out}}}(t)=0$, which from
eq. 3 implies ${{S}_{\text{\text in}}}(t)=\sqrt{2{{\gamma
    }_{s}}}S(t)$. Then using the bad-cavity approximation for $S$, as
given by eq. 4, leads straightforwardly to ${{\partial
  }_{t}}C(t)=K(t)C(t)$ and ${{S}_{\text in}}(t)={{\mu
  }^{*}}\sqrt{2K}C(t)$, where $K(t)={{f}_{s}}|\Omega (t){{|}^{2}}$
and $\mu =i\exp \{i\arg [\Omega (t)]\}$. From these, one can derive
the design equation for $K(t)$:

\begin{equation}
  \left( \frac{{{\partial }_{t}}K}{2K} \right)+K(t)=\frac{{{\partial }_{t}}{{S}_{\text in}}(t)}{{{S}_{\text in}}(t)}.\label{eq8}
\end{equation}

An equation of this form also appears in the context of optical
storage in cavity-enhanced atomic quantum memories, where it is called
the ``impedance matching condition.'' \cite{Cirac:04ax}. The resulting
solution for the control field for optimal storage is (See the
Appendix.)

\begin{equation}
  {{\Omega }_{\text opt}}(t)={{e}^{i\theta }}\,{{e}^{-i\arg [{{S}_{\text{\text in}}}(t)]}}\sqrt{\frac{{{S}_{\text{\text in}}}{{(t)}^{2}}}{q+2{{f}_{s}}\int\limits_{{{t}_{0}}}^{t}{{{S}_{\text{\text in}}}}{{({t}')}^{2}}d{t}'}},\label{eq9}
\end{equation}

\noindent where $\theta $ is an arbitrary phase and
$q={{S}_{\text{\text in}}}{{({{t}_{0}})}^{2}}/|\Omega
({{t}_{0}}){{|}^{2}}$, which for numerical purposes is a vanishingly
small parameter if the arbitrary initial time ${{t}_{0}}$ is taken to
be well before the input signal begins rising from zero value.

To illustrate and test this design prediction, consider as a target
input signal any of the orthogonal Hermite Gaussians,
$H{{G}_{n}}(t)={{H}_{n}}(t){{e}^{-{{t}^{2}}/2}}/\sqrt{{{2}^{n}}{{\pi
    }^{1/2}}n!}$. We numerically solve eqs. 1 and 2 using eq. \ref{eq9} as the
control field, and plot the results in fig. 4, where Figs. 4a and 4b
show the results for ${{S}_{\text in}}(t)=H{{G}_{0}}(t)$ and
$H{{G}_{1}}(t)$, respectively. The value of the control strength
parameter $\alpha $ is again optimized to the value 5.5. The shapes of
the control fields before the signal begins turning on are arbitrary,
and set by the value ${{10}^{-7}}$ of the parameter $q$. The
unconverted signal energy in case (a) is 0.004, and in case (b) is
0.015, showing excellent conversion and trapping of the targeted input
TMs.

\begin{figure}[bht]
  \centering
  \includegraphics[width=\linewidth]{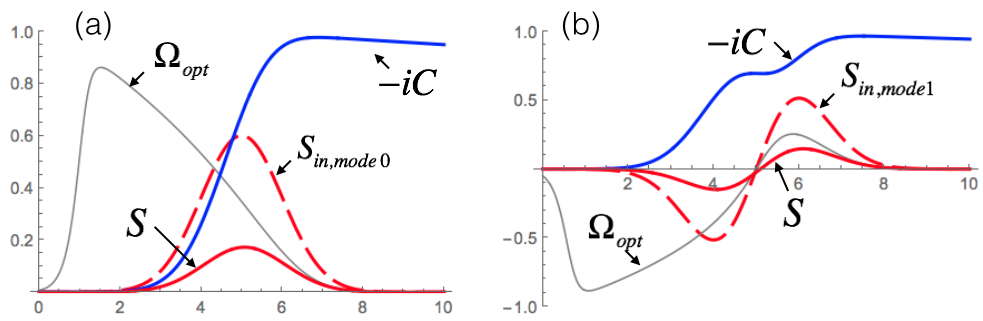}
  \caption{Illustrating the effectiveness of the control field
    ${{\Omega }_{opt}}(t)$ to efficiently convert and store the
    targeted ``red'' input
    mode ${{S}_{\text in}}(t)$. (a) ${{S}_{\text in}}(t)=H{{G}_{0}}(t)$,
    (b) ${{S}_{\text in}}(t)=H{{G}_{1}}(t)$. In both cases, using the
    designed control field drives the converted cavity mode amplitude
    $-iC$ to near its maximum possible value of 1.0. In both cases:
    $\alpha =5.5,\ {{\gamma }_{s}}=10.1,\ {{\gamma
      }_{c}}=0.01,\ q={{10}^{-7}}$.}
  \label{fig04}
\end{figure}

If we choose the unit time scale in the simulations as 100 ps, the
duration of the control pulse in Fig. 2 is 166 ps. Assuming a
round-trip time 15 ps and group velocity c/2 gives a cavity round-trip
length 225 $\mu m$. The cavity leakage parameters (${{\gamma
  }_{s}}=10.1$, ${{\gamma }_{c}}=0.01$) corresponding to rates
$1.01\times {{10}^{11}}{{s}^{-1}}$and $1.0\times {{10}^{8}}{{s}^{-1}}$
(and cavity-field lifetimes 10 ps and 10 ns) respectively. For carrier
wavelengths 1550 nm and 775 nm for $S$ and $C$ modes, respectively,
the ``dichroic'' cavity quality factors needed are ${{Q}_{s}}={{\omega
  }_{s}}/2{{\gamma }_{s}}=6020$ and ${{Q}_{c}}={{\omega
  }_{c}}/2{{\gamma }_{c}}=1.2\times {{10}^{7}}$. Finally, the internal
dissipative losses need to be much smaller than ${{\gamma
  }_{s}},\ {{\gamma }_{c}}$. These values, along with the needed
control power, are within range of achievable values for
whispering-gallery resonators or planar-waveguide micro-rings coupled
evanescently to an external waveguide.

In summary, the newly proposed scheme offers several functionalities
crucial for quantum information science. It can multiplex and
demultiplex orthogonal temporal modes of light with high TM
discrimination and efficiency. It is reconfigurable in real time to
target any chosen TM by altering the shape of the control field that
drives the sum-frequency generation. The efficiency of TM
demultiplexing is tunable in real time by altering the intensity of
the control field, giving the possibility to create and measure
single-photon states that are superpositions of two time-bin
states. It can be used as a short-time all-optical memory, the storage
time being limited by cavity Q and losses. And it can be used to
reshape optical pulses via the read-in, read-out process.

The vast majority of time-stationary optical processes satisfy the
Fourier constraint, $\Delta {{\omega }_{P}}\Delta {{t}_{P}}\approx
2\pi $, where $\Delta {{\omega }_{P}}$ and $\Delta {{t}_{P}}$ are the
bandwidth and processing (interaction, or read-out) time of the
process. (Rare exceptions may occur in systems lacking Lorentz
(time-reversal) reciprocity \cite{Tsakmakidis1260}.) Our system, being
time-nonstationary, has the useful property that both the bandwidth
and the storage time, while not being bound by $\Delta {{\omega
  }_{P}}\Delta {{t}_{P}}\approx 2\pi $, are tunable in real time. The
read-in bandwidth $\Delta {{\omega }_{P}}$ for the input channel is set
by the shape of the control field $\Omega (t)$, while the read-out
time $\Delta {{t}_{P}}$ in the output channel is set either by the
natural decay time of the narrow $C$ mode resonance (which is much
narrower than that of the input $S$ mode resonance) or by the duration
of the outgoing red signal pulse in cases where a read control pulse
is employed.

Acknowledgements: We thank Hailin Wang for helpful discussions. DVR
and MGR were supported by NSF grant no. 1521466.

\section*{Appendix}

The solution of eq. \ref{eq8} for the optimum coupling function is

\begin{equation}
  K(t)=\frac{K({{t}_{0}}){{S}_{\text{\text in}}}{{(t)}^{2}}}{{{S}_{\text{\text in}}}{{({{t}_{0}})}^{2}}+2K({{t}_{0}})\int\limits_{{{t}_{0}}}^{t}{{{S}_{\text{\text in}}}}{{({t}')}^{2}}d{t}'}
  \end{equation}

Using 
	$K(t)={{f}_{s}}|\Omega (t){{|}^{2}}$
 and 
	$\mu =i\exp \left\{ i\arg [\Omega (t)] \right\}$
 gives

 \begin{equation}
   \mu \sqrt{2K(t)}=i\,{{e}^{i\arg [\Omega (t)]}}\sqrt{2{{f}_{s}}}\sqrt{|\Omega (t){{|}^{2}}}=i\sqrt{2{{f}_{s}}}\Omega (t)
   \end{equation}
	 
Therefore

\begin{align}
  & \mu \sqrt{2K(t)}=i\,{{e}^{i\arg [\Omega (t)]}}\sqrt{2{{f}_{s}}}\sqrt{|\Omega (t){{|}^{2}}}=i\sqrt{2{{f}_{s}}}\Omega (t) \\ 
 & {{\Omega }_{opt}}(t)=\frac{\mu \sqrt{2K(t)}}{i\sqrt{2{{f}_{s}}}}={{e}^{i\arg [\Omega (t)]}}\sqrt{\frac{K(t)}{{{f}_{s}}}}={{e}^{i\arg [\Omega (t)]}}\sqrt{\frac{{{S}_{\text{\text in}}}{{(t)}^{2}}}{{{S}_{\text{\text in}}}{{({{t}_{0}})}^{2}}{{f}_{s}}/K({{t}_{0}})+2{{f}_{s}}\int\limits_{{{t}_{0}}}^{t}{{{S}_{\text{\text in}}}}{{({t}')}^{2}}d{t}'}} \\ 
 & {{\Omega }_{opt}}(t)={{e}^{i\arg [\Omega (t)]}}\sqrt{\frac{{{S}_{\text{\text in}}}{{(t)}^{2}}}{{{S}_{\text{\text in}}}{{({{t}_{0}})}^{2}}/|\Omega ({{t}_{0}}){{|}^{2}}+2{{f}_{s}}\int\limits_{{{t}_{0}}}^{t}{{{S}_{\text{\text in}}}}{{({t}')}^{2}}d{t}'}}
\end{align}

For the optimum input TM case, the time-derivative of $C(t)$ in eq. 5
should always have the same phase, as we require that $|C(t)|$ grow
monotonically. For the second term in the right-hand side of eq. 5 to
have constant phase for arbitrary input $S_{\text in}(t)$, we need
$arg[{{\Omega }_{opt}}(t)]=\theta -\arg [{{S}_{\text{\text in}}}(t)],$
leading posthaste to eq. \ref{eq9}.

\begin{align}
  \mu \sqrt{2K(t)}&=i\sqrt{2{{f}_{s}}}{{e}^{i\arg [\Omega (t)]}}\sqrt{|\Omega (t){{|}^{2}}}=i\sqrt{2{{f}_{s}}}\Omega (t) \\ 
{{\Omega }_{opt}}(t)&={{e}^{i\theta }}{{e}^{-i\arg [{{S}_{\text{\text in}}}(t)]}}\sqrt{\frac{{{S}_{\text{\text in}}}{{(t)}^{2}}}{q+2{{f}_{s}}\int\limits_{{{t}_{0}}}^{t}{{{S}_{\text{\text in}}}}{{({t}')}^{2}}d{t}'}}
\end{align}


\begin{thebibliography}{10}
\newcommand{\enquote}[1]{``#1''}

\bibitem{ptd12}
M.~G. Raymer and K.~Srinivasan, \enquote{{Manipulating the color and shape of
  single photons},} Physics Today \textbf{65}, 32 (2012).

\bibitem{Humphreys:2014ce}
P.~C. Humphreys, W.~S. Kolthammer, J.~Nunn, M.~Barbieri, A.~Datta, and I.~A.
  Walmsley, \enquote{{Continuous-Variable Quantum Computing in Optical
  Time-Frequency Modes Using Quantum Memories},} Physical Review Letters
  \textbf{113}, 130502 (2014).

\bibitem{nunn13}
J.~Nunn, L.~J. Wright, C.~S\"{o}ller, L.~Zhang, I.~A. Walmsley, and B.~J.
  Smith, \enquote{Large-alphabet time-frequency entangled quantum key
  distribution by means of time-to-frequency conversion,} Opt. Express
  \textbf{21}, 15959--15973 (2013).

\bibitem{Wright}
L.~J. Wright, M.~Karpi\ifmmode~\acute{n}\else \'{n}\fi{}ski, C.~S\"oller, and
  B.~J. Smith, \enquote{Spectral shearing of quantum light pulses by
  electro-optic phase modulation,} Phys. Rev. Lett. \textbf{118}, 023601
  (2017).

\bibitem{Davis:17}
A.~O.~C. Davis, P.~M. Saulnier, M.~Karpi\'{n}ski, and B.~J. Smith,
  \enquote{Pulsed single-photon spectrometer by frequency-to-time mapping using
  chirped fiber bragg gratings,} Opt. Express \textbf{25}, 12804--12811 (2017).

\bibitem{Smith:2007njp}
B.~J. Smith and M.~G. Raymer, \enquote{Photon wave functions, wave-packet
  quantization of light, and coherence theory,} New Journal of Physics
  \textbf{9}, 414 (2007).

\bibitem{Brecht:15prx}
B.~Brecht, D.~V. Reddy, C.~Silberhorn, and M.~G. Raymer, \enquote{Photon
  temporal modes: A complete framework for quantum information science,} Phys.
  Rev. X \textbf{5}, 041017 (2015).

\bibitem{EcksteinA:2011vg}
A.~Eckstein, B.~Brecht, and C.~Silberhorn, \enquote{{A quantum pulse gate based
  on spectrally engineered sum frequency generation},} Optics Express
  \textbf{19}, 13370 (2011).

\bibitem{Brecht:2011hz}
B.~Brecht, A.~Eckstein, A.~Christ, H.~Suche, and C.~Silberhorn, \enquote{{From
  quantum pulse gate to quantum pulse shaper{\textemdash}engineered frequency
  conversion in nonlinear optical waveguides},} New Journal of Physics
  \textbf{13}, 065029 (2011).

\bibitem{Reddy:2013ip}
D.~V. Reddy, M.~G. Raymer, C.~J. McKinstrie, L.~Mejling, and K.~Rottwitt,
  \enquote{{Temporal mode selectivity by frequency conversion in second-order
  nonlinear optical waveguides},} Optics Express \textbf{21}, 13840--13863
  (2013).

\bibitem{Ansari:16ax}
V.~Ansari, M.~Allgaier, L.~Sasoni, B.~Brecht, J.~Roslund, N.~Treps, G.~Harder,
  and C.~Silberhorn, \enquote{Temporal-mode tomography of single photons,}
  arXiv:1607.03001v1  (2016).

\bibitem{Manurkar:16}
P.~Manurkar, N.~Jain, M.~Silver, Y.-P. Huang, C.~Langrock, M.~M. Fejer,
  P.~Kumar, and G.~S. Kanter, \enquote{Multidimensional mode-separable
  frequency conversion for high-speed quantum communication,} Optica
  \textbf{3}, 1300--1307 (2016).

\bibitem{Reddy:17}
D.~V. Reddy and M.~G. Raymer, \enquote{Engineering temporal-mode-selective
  frequency conversion in nonlinear optical waveguides: from theory to
  experiment,} Opt. Express \textbf{25}, 12952--12966 (2017).

\bibitem{Reddy:2014bt}
D.~V. Reddy, M.~G. Raymer, and C.~J. McKinstrie, \enquote{{Efficient sorting of
  quantum-optical wave packets by temporal-mode interferometry},} Optics
  Letters \textbf{39}, 2924--2927 (2014).

\bibitem{Reddy:15pra}
D.~V. Reddy, M.~G. Raymer, and C.~J. McKinstrie, \enquote{Sorting photon wave
  packets using temporal-mode interferometry based on multiple-stage quantum
  frequency conversion,} Phys. Rev. A \textbf{91}, 012323 (2015).

\bibitem{Kobayashi:17ax}
T.~Kobayashi, D.~Yamazaki, K.~Matsuki, R.~Ikuta, S.~Miki, T.~Yamashita,
  H.~Terai, T.~Yamamoto, M.~Kaoshi, and N.~Imoto, \enquote{{Mach-Zehnder
  interferometer using frequency-domain beamsplitter},} arXiv:1703.08114
  (2017).

\bibitem{Gaeta:15ax}
S.~Ramelow, A.~Farsi, S.~Clemmen, D.~Orquiza, K.~Luke, M.~Lipson, and A.~L.
  Gaeta, \enquote{Silicon-nitride platform for narrowband entangled photon
  generation,} arXiv:1508.04358  (2015).

\bibitem{Quesada:16}
N.~Quesada and J.~E. Sipe, \enquote{High efficiency in mode-selective frequency
  conversion,} Opt. Lett. \textbf{41}, 364--367 (2016).

\bibitem{Zheng:01}
Z.~Zheng and A.~Weiner, \enquote{Coherent control of second harmonic generation
  using spectrally phase coded femtosecond waveforms,} Chemical Physics
  \textbf{267}, 161 -- 171 (2001).

\bibitem{Zheng:02}
Z.~Zheng, A.~M. Weiner, K.~R. Parameswaran, M.-H. Chou, and M.~M. Fejer,
  \enquote{Femtosecond second-harmonic generation in periodically poled lithium
  niobate waveguides with simultaneous strong pump depletion and group-velocity
  walk-off,} J. Opt. Soc. Am. B \textbf{19}, 839--848 (2002).

\bibitem{Cirac:04ax}
J.~I. Cirac, L.~M. Duan, and P.~Z{\"o}ller, \enquote{Quantum optical
  implementation of quantum information processing,} arXiv:quant-ph/0405030
  (2004).

\bibitem{Bao:12}
X.-H. Bao, A.~Reingruber, P.~Dietrich, J.~Rui, A.~Duck, T.~Strassel, L.~Li,
  N.-L. Liu, B.~Zhao, and J.-W. Pan, \enquote{Efficient and long-lived quantum
  memory with cold atoms inside a ring cavity,} Nat. Phys. \textbf{8}, 517--521
  (2012).

\bibitem{Nunn:16ax}
J.~Nunn, S.~Thomas, J.~H.~D. Munns, K.~T. Kaczmarek, C.~Qiu, A.~Feizpour,
  E.~Poem, B.~Brecht, D.~J. Saunders, P.~M. Ledingham, D.~V. Reddy, M.~G.
  Raymer, and I.~A. Walmsley, \enquote{Theory of noise suppression in
  $\lambda$-type quantum memories by means of a cavity,} arXiv:1601.00157v1
  (2016).

\bibitem{Strekalov}
D.~V. Strekalov, A.~S. Kowligy, Y.-P. Huang, and P.~Kumar, \enquote{Optical
  sum-frequency generation in a whispering-gallery-mode resonator,} New Journal
  of Physics \textbf{16}, 053025 (2014).

\bibitem{Li}
Q.~Li, M.~Davanco, and K.~Srinivasan, \enquote{Efficient and low noise
  single-photon-level frequency conversion interfaces using si3n4 microrings,}
  in \enquote{2016 Progress in Electromagnetic Research Symposium (PIERS),}
  (2016), pp. 2574--2574.

\bibitem{Guo}
X.~Guo, C.-L. Zou, H.~Jung, and H.~X. Tang, \enquote{On-chip strong coupling
  and efficient frequency conversion between telecom and visible optical
  modes,} Phys. Rev. Lett. \textbf{117}, 123902 (2016).

\bibitem{kumarcav}
D.~V. Strekalov, C.~Marquardt, A.~B. Matsko, H.~G.~L. Schwefel, and G.~Leuchs,
  \enquote{Nonlinear and quantum optics with whispering gallery resonators,}
  Journal of Optics \textbf{18}, 123002 (2016).

\bibitem{Burnham1969}
D.~Burnham and R.~Chiao, \enquote{{Coherent Resonance Fluorescence Excited by
  Short Light Pulses},} Physical Review \textbf{188}, 667--675 (1969).

\bibitem{Collett}
M.~J. Collett and C.~W. Gardiner, \enquote{Squeezing of intracavity and
  traveling-wave light fields produced in parametric amplification,} Phys. Rev.
  A \textbf{30}, 1386--1391 (1984).

\bibitem{Raymercav}
M.~G. Raymer and C.~J. McKinstrie, \enquote{Quantum input-output theory for
  optical cavities with arbitrary coupling strength: Application to two-photon
  wave-packet shaping,} Phys. Rev. A \textbf{88}, 043819 (2013).

\bibitem{kumarcav2}
Y.-Z. Sun, Y.-P. Huang, and P.~Kumar, \enquote{Photonic nonlinearities via
  quantum zeno blockade,} Phys. Rev. Lett. \textbf{110}, 223901 (2013).

\bibitem{shen}
J.-T. Shen and S.~Fan, \enquote{Theory of single-photon transport in a
  single-mode waveguide. ii. coupling to a whispering-gallery resonator
  containing a two-level atom,} Phys. Rev. A \textbf{79}, 023838 (2009).

\bibitem{Mirza:13}
I.~M. Mirza, S.~J. van Enk, and H.~J. Kimble, \enquote{Single-photon
  time-dependent spectra in coupled cavity arrays,} J. Opt. Soc. Am. B
  \textbf{30}, 2640--2649 (2013).

\bibitem{Tsakmakidis1260}
K.~L. Tsakmakidis, L.~Shen, S.~A. Schulz, X.~Zheng, J.~Upham, X.~Deng,
  H.~Altug, A.~F. Vakakis, and R.~W. Boyd, \enquote{{Breaking Lorentz
  reciprocity to overcome the time-bandwidth limit in physics and
  engineering},} Science \textbf{356}, 1260--1264 (2017).

\end{thebibliography}
\end{document}